\documentclass[Afour,sagev,times]{sagej}

\usepackage{moreverb,url}

\usepackage[colorlinks,bookmarksopen,bookmarksnumbered,citecolor=red,urlcolor=red]{hyperref}

\usepackage{amsmath,amsthm,amsfonts,amssymb,amscd}
\usepackage{enumerate}
\usepackage{fancyhdr}
\usepackage{mathrsfs}
\usepackage{xcolor}
\usepackage{graphicx}
\usepackage{listings}
\usepackage{hyperref}
\usepackage{orcidlink}
\usepackage{float}
\usepackage{enumitem,kantlipsum}
\usepackage{tabularx}
\usepackage{caption}
\providecommand{\abs}[1]{\lvert#1\rvert}
\providecommand{\norm}[1]{\lVert#1\rVert}

\newtheorem{theorem}{Theorem}

\newtheorem{remark}{Remark}
\newtheorem{lemma}{Lemma}

\newtheorem{definition}{Definition}

\newcommand\BibTeX{{\rmfamily B\kern-.05em \textsc{i\kern-.025em b}\kern-.08em
T\kern-.1667em\lower.7ex\hbox{E}\kern-.125emX}}

\begin{document}

\title{Stabilizing unstable periodic orbit of unknown fractional-order systems via adaptive delayed feedback control}

\author{Bahram Yaghooti \orcidlink{0000-0001-9646-9687} \affilnum{1,}\affilnum{*}, Kaveh Safavigerdini \orcidlink{0000-0003-4904-0161} \affilnum{2,}\affilnum{*}, Reza Hajiloo \affilnum{3} and Hassan Salarieh \orcidlink{0000-0002-0604-5731} \affilnum{4}}

\affiliation{\affilnum{1}Department of Electrical and Systems Engineering, Washington University in St. Louis, St. Louis, MO 63130, USA. \\
\affilnum{2}Department of Electrical and Computer Engineering, University of Missouri, Columbia, MO 65211, USA. \\
\affilnum{3}Department of Mechanical and Mechatronics Engineering, University of Waterloo, Waterloo, ON N2L3G1, Canada. \\
\affilnum{4}Department of Mechanical Engineering, Sharif University of Technology, Tehran, Tehran, Iran.\\
\affilnum{*} These authors contributed equally: Bahram Yaghooti and Kaveh Safavigerdini.
}

\corrauth{Kaveh Safavigerdini, Department of Electrical and Computer Engineering, University of Missouri, Columbia, MO, 65211, USA.}

\email{ksgh2@umsystem.edu}

\begin{abstract}
This article presents an adaptive nonlinear delayed feedback control scheme for stabilizing the unstable periodic orbit of unknown fractional-order chaotic systems. The proposed control framework uses the Lyapunov approach and sliding mode control technique to guarantee that the closed-loop system is asymptotically stable on a periodic trajectory sufficiently close to the unstable periodic orbit of the system. The proposed method has two significant advantages. First, it employs a direct adaptive control method, making it easy to implement this method on systems with unknown parameters. Second, the framework requires only the period of the unstable periodic orbit. The robustness of the closed-loop system against system uncertainties and external disturbances with unknown bounds is guaranteed. Simulations on fractional-order duffing and gyro systems are used to illustrate the effectiveness of the theoretical results. The simulation results demonstrate that our approach outperforms the previously developed linear feedback control method for stabilizing unstable periodic orbits in fractional-order chaotic systems, particularly in reducing steady-state error and achieving faster convergence of tracking error. 
\end{abstract}


\maketitle

\section{Introduction}\label{sec:introduction}
Fractional Calculus is the generalization of integrals and derivatives to non-integer orders. It has been used in several fields, for example, soft robotics \cite{mousa2020biohybrid, azimirad2022consecutive}, vibration \cite{sahoo2019active}, electromagnetic problems \cite{engheta1996fractional}, power grid \cite{shalalfeh2018modeling}, piezoelectric actuators \cite{yu2021extended}, and drug delivery systems \cite{yaghooti2020constrained}. There are several reasons that fractional-order (FO) systems have been employed in control theory and dynamical systems, for instance, their accuracy in modeling dynamical systems, larger stability areas, and capacity to satisfy more control criteria \cite{yaghooti2023inferring,azar2017fractional}. During the last few years, numerous control schemes have been developed for FO systems. Shahvali et al. studied the distributed adaptive dynamic event-based consensus control for nonlinear uncertain multi-agent systems \cite{shahvali2022distributed}. Zamani et al. addressed the fixed-time formation problem in FO multi-agent systems \cite{zamani2022formation}. Rezaei Lori et al. developed a sliding mode control method to control position of a quadrotor in the presence of external disturbances \cite{lori2021transportation,amiri2020fuzzy}. Ayten et al. presented the speed and direction angle control of a wheeled mobile robot based on a FO adaptive model-based PID-type sliding mode control technique \cite{ayten2019implementation}. Machado et al. addressed the modeling and dynamical analysis of soccer teams by two modeling perspectives which are adopted based on the concepts of fractional calculus \cite{machado2017mathematical}. Shahvali et al. proposed an adaptive fault compensation control for nonlinear uncertain FO systems \cite{shahvali2021adaptive}. Huong et al. studied the problem of global asymptotic stability analysis, and mixed $H_\infty$ and passive control for a class of control FO nonlinear systems using the Lyapunov direct method \cite{huong2020mixed}. Essa et al. represented the application of fractional order controllers on an experimental and simulation model of a hydraulic servo system \cite{essa2017application}.  Wang et al. investigated an adaptive fractional‐order nonsingular terminal sliding mode control scheme based on time delay estimation \cite{wang2019practical}. Yaghooti et al. proposed an adaptive FO PID control for FO nonlinear systems \cite{yaghooti2018robust}. Furthermore, all practical control systems are subject to operational constraints, including restricted dimensions and limited control capacity. Solutions to these issues have been proposed in the form of Model Predictive Control and Explicit Reference Governor \cite{askari2022sampling, askari2021nonlinear, azimirad2020optimizing, nicotra2018explicit}. Similar to integer-order systems, FO systems can also be subject to these operational constraints. Consequently, these constrained control schemes have been developed and adapted for use with FO systems. \cite{yaghooti2020constrained}

The research area of chaotic dynamics and control of FO dynamical systems has gained significant attention and popularity in recent years. Chaos is an intriguing phenomenon of FO systems which can endanger many systems. Nonlinear FO systems of orders lower than three are able to generate chaotic attractors  \cite{aghababa2013rich,yaghooti2020adaptive}. Moreover, it is demonstrated that UPO found in these systems can be served as a generalization of the integer-order case and potentially provides more accurate system modeling for a number of applications in robotics \cite{farid2021finite,vu2021iterative} and optimal motion planning \cite{oshima2019spatial,he2022homotopy}. Some control techniques have been applied for stabilizing the UPO of FO chaotic systems such as linear feedback control. Rahimi et al. represented techniques for finding unstable periodic orbits in chaotic fractional order systems and for the stabilization of founded UPOs \cite{rahimi2012stabilizing}. Sadeghian et al. illustrated a method to apply feedback of measured states using the period of fixed points (in discrete systems) and periodic orbits (in continuous systems) which there is no need for information for fixed point and periodic orbits \cite{sadeghian2011control}. Naceri et al. introduced a control algorithm based on the delayed feedback control method to stabilize a fractional order system on an unstable fixed point \cite{naceri2008prediction}.
Zheng et al. presented the stabilization of a fractional order chaotic system on its original equilibrium point using the Takagi–Sugeno (T–S) fuzzy models and prediction-based feedback controls \cite{zheng2015fuzzy}. Shahvali et al. created an adaptive neural network (NN) backstepping control method for the nonlinear double-integrator FO systems \cite{9615365}. Rabah et al. studied the behavior of the fractional Lü system and chaos suppression using a fractional order proportional integral derivative controller \cite{rabah2015state}. Danca et al. investigated a class of nonlinear impulsive Caputo differential equations of fractional order, which models chaotic systems \cite{danca2017impulsive}. Layeghi et al. designed a fuzzy adaptive sliding mode scheme to stabilize the unstable periodic orbits of chaotic systems by using a combination of fuzzy identification and sliding mode control \cite{layeghi2008stabilizing}. 

The control schemes previously discussed exhibit two significant limitations. Firstly, they are developed based on systems whose dynamics are fully known. However, in many cases, it is challenging to analytically determine system parameters due to uncertainties and external disturbances. Secondly, these algorithms necessitate complete knowledge of the unstable periodic orbit's trajectory, which may not always be feasible to obtain. 

This article introduces a new algorithm for stabilizing unstable periodic orbits in unknown FO chaotic systems. The proposed algorithm addresses the limitations of previous techniques by utilizing a direct adaptive control method that can be applied to systems with unknown parameters without the need for prior knowledge of system dynamics. Additionally, the proposed framework only requires knowledge of the UPO's period, eliminating the need for complete information about the UPO's trajectory as required by previously developed methods in the literature. It is assumed that the system is subject to uncertainties and external disturbances with unknown bounds. Therefore, an adaptive framework is designed such that the upper bounds of uncertainties and disturbances are estimated while stabilizing the system’s UPO. To do so, this method utilizes an adaptive nonlinear delayed feedback control along with a sliding mode control technique. In order to cope with the external disturbances and system uncertainties, an appropriate sliding surface is utilized to derive adaptation laws, resulting in a stable closed-loop system.

The rest of the article is organized as follows: Section \nameref{sec:preliminary_concepts}
illustrates a succinct review of the fundamental principles of fractional calculus. Section \nameref{sec:problem_statement} presents the problem formulation. In Section \nameref{sec:control_design}, an adaptive sliding mode control method is proposed. The parameters are updated using an adaptation mechanism. In Section \nameref{sec:simulation_results}, the developed control method is implemented to stabilize UPOs of FO chaotic Gyro and Duffing Systems. Numerical simulation results support the efficacy of the suggested framework. The final section presents a succinct conclusion.

\section{Preliminaries}\label{sec:preliminary_concepts}
The theory of non-integer order derivatives was first introduced by Leibnitz and subsequently, several definitions were proposed. In this article, we use Caputo's definition. The purpose of this section is to give a brief overview of key concepts and stability theorems that are fundamental to the study of FO systems.

\begin{definition} The fractional integral of order $\alpha\in\mathbb{R}$ is defined as \cite{podlubny1998fractional}
\begin{align}
    _{0}I^{\alpha}_{t}f(t)= 
        {\frac{1}{\Gamma(\alpha)} \int_{0}^{t} (t-\tau)^{\alpha-1}f(\tau) d\tau},
\end{align}
where $\Gamma(\cdot)$ stands for the standard Gamma function, such that $\Gamma(x) = \int_{0}^{\infty} t^{x-1}e^{-t} dt$.
\end{definition}

\begin{definition} The Caputo fractional derivative of order $\alpha\in\mathbb{R}$ is defined as \cite{podlubny1998fractional} 
\begin{equation}\label{eq:caputo_definition}
    {}^{C}_{0}D^{\alpha}_{t}f(t)= 
    \begin{cases}
        \frac{1}{\Gamma(p-\alpha)} \int_{0}^{t} \frac{f^{(p)}(\tau)}{(t-\tau)^{\alpha+1-p}} d\tau; & {p-1<\alpha<p}\\
        \frac{d^{p}}{dt^{p}}f(t); & {\alpha=p}\\
    \end{cases},
\end{equation}
where $p$ is an integer number such that $p-1<\alpha<p$.
\end{definition}

\textit{Notation}. For the sake of concision, we will use $D^\alpha_t$ and $I^{\alpha}_t$ to denote the operators ${}^{C}_{0}D^{\alpha}_{t}$ and $_0I^{\alpha}_t$.

\begin{theorem}\label{thm:linear_system_stability}
Let us define a commensurate certain FO linear system as \cite{matignon1996stability}
\begin{align}\label{eq:general_linear_system}
    D_t^{\alpha}x = Ax,
\end{align}
where $x\in\mathbb{R}^n$ represents the state vector of the system; $A\in\mathbb{R}^{n\times n}$ is the state matrix; and $\alpha\in\mathbb{R}$ belongs to an interval $(0,2)$ and stands for the order of fractional derivatives. The asymptotic stability of the system represented by equation \eqref{eq:general_linear_system} can be determined by evaluating the satisfaction of the following criteria:
\begin{align}\label{eq:linear_stability}
    \abs{\arg(\lambda_i)}>\alpha \frac{\pi}{2}\text{ }; \quad 1\leq i \leq n,
\end{align}
where $\lambda_i$, $1\leq i \leq n$, denotes the eigenvalues of matrix $A$.
\end{theorem}

By using the Lyapunov direct method, the asymptotic stability of FO nonlinear systems can be studied. In the following theorem, the Lyapunov direct method is extended to the case of FO systems, which leads to the Mittag–Leffler stability \cite{li2010stability}.

\begin{theorem} By using the Caputo derivative, a FO nonlinear system is defined as \cite{li2010stability} 
\begin{align}\label{eq:general_nonlinear_system}
    D^\alpha _t x=f(t,x),
\end{align}
where $x\in\mathbb{R}^n$ presets the state vector of the system; $\alpha\in\mathbb{R}$ is the order of fractional derivatives of the system. Let $x=0$ represent the system \eqref{eq:general_nonlinear_system} equilibrium point, 
and $V(t,x(t))$ be a continuously differentiable function and a candidate for the Lyapunov function, and also $\zeta_i$ ($i=1,2,3$) be class-$\kappa$ functions such that 
\begin{align}
    & \zeta_{1}(\left\|x\right\|){\leq} V(t,x(t)){\leq}\zeta_{2}(\left\|x\right\|), \\
    & D^{\nu}_{t}V(t,x(t))  {\leq}-\zeta_{3}(\left\|x\right\|),
\end{align}
where ${\nu\in(0,1)}$. So, the FO system of \eqref{eq:general_nonlinear_system} is asymptotically stable.
\end{theorem}
In what follows, we introduce some lemmas which will be used to prove the stability of our proposed algorithm.
\begin{lemma}\label{lemma:derivativeLF} 
Let ${x(t)\in\mathbb{R}^n}$ represent a vector of continuously differentiable functions. Afterward, for all of ${t\geq0}$ it continues to hold \cite{duarte2015using}

\begin{align}
    {D^\alpha_t(x^\top Px)}\leq{\left(D_t^\alpha x\right)^\top Px+x^\top PD^\alpha_tx}; \quad \forall\alpha\in(0,1),
\end{align}
where ${P\in\mathbb{R}^{n\times n}}$ states a symmetric positive definite matrix.
\end{lemma}
Barbalat’s Lemma is developed for FO nonlinear systems as follows \cite{zhang2017new}:
\begin{lemma}\label{lemma:barbalat} 
Let $\psi:\mathbb{R}\to\mathbb{R}$ represent a uniformly continuous function on $[0,\infty)$. 
Suppose $q$ and $N$ are two positive constants such that $I_t^{\alpha}\abs{\psi}^{q}\leq N$ for all $t>0$ with $\alpha\in(0,1)$. Then \cite{zhang2017new}
\begin{align}
    \lim_{t\to\infty} \psi(t) = 0.
\end{align}
\end{lemma}

\section{Problem Statement}\label{sec:problem_statement}
Consider the following control-affine FO chaotic system
\begin{align}\label{eq:nonlinear_system}
\begin{cases}
D^{\alpha}_{t}x_i =x_{i+1}; \qquad i=1, \dots, n-1\\
D^{\alpha}_{t}x_n = f(t,x) + F^\top(t,x)\theta + d(t) + g(t,x)u(t),
\end{cases}
\end{align}
where $0<\alpha<1$ represents the fractional derivative order of the system, $x=[x_1,\dots,x_n]^\top\in\mathbb{R}^n$ is the system state vector, $f(\cdot, \cdot)$ and $g(\cdot, \cdot)$ are nonlinear and continuously differentiable functions. It is assumed that $g(\cdot, \cdot)\neq 0$, and that $F^\top(t,x)\theta$ represents the system uncertainties, where $F(\cdot, \cdot)\in\mathbb{R}^m$ is known and $\theta\in\mathbb{R}^m$ is a vector of unknown parameters. $d(\cdot)\in\mathbb{R}$ represents the external disturbance, and it is assumed that the disturbance is bounded ($\abs{d(\cdot)}\leq k < \infty$), but the value of boundary ($k>0$) is unknown. The system input is represented by $u(t)\in\mathbb{R}$ and it is assumed that the system exhibits chaotic behavior for $u=0$. The function $f(t,x) + F^\top(t,x)\theta$ is $t$-periodic with period $T$, and without any control or disturbance, the system \eqref{eq:nonlinear_system} exhibits an unstable periodic response with a known period of $T$.

The primary objective of this study is to design and implement a robust adaptive control framework that effectively makes the chaotic system track a periodic orbit that is approximate to one of the system's UPOs. To achieve this, an adaptive control approach is employed, as the exact plant function is not fully known. The proposed adaptive laws are designed to ensure the stability of the closed-loop control system, while the use of sliding mode techniques provides robustness against uncertainties and external disturbances of unknown magnitudes. This research aims to provide a novel solution to the challenging problem of chaotic systems and contribute to the field of control engineering.

\section{Stability Analysis and Control Scheme Design}\label{sec:control_design}
This section presents the design of a nonlinear delayed feedback control framework for the FO system defined by equation \eqref{eq:nonlinear_system}. A delayed state is defined as $\Tilde{x}(t) = x(t-T)$, and it is substituted into the system dynamics given by equation \eqref{eq:nonlinear_system} to obtain the delayed state dynamics as follows:
\begin{align}\label{eq:nonlinear_system_tilde}
\begin{cases}
D^{\alpha}_{t}\Tilde{x}_i  = \Tilde{x}_{i+1}; \qquad i = 1, \dots, n-1\\
D^{\alpha}_{t}\Tilde{x}_n  = f(t-T,\Tilde{x})+F^\top(t-T,\Tilde{x})\Tilde{\theta} \\
\qquad\qquad + \; \Tilde{d}(t) + g(t-T,\Tilde{x})\Tilde{u}(t)
\end{cases},
\end{align}
where $\Tilde{u}(t) = u(t-T)$, $\Tilde{d}(t) = d(t-T)$, and $\Tilde{\theta} = \theta(t-T)$. To obtain the closed-loop system's error dynamics, one can subtract \eqref{eq:nonlinear_system_tilde} from \eqref{eq:nonlinear_system}. 
\begin{align}\label{eq:error_1}
\begin{cases}
D^{\alpha}_{t}x_i-D^{\alpha}_{t}\Tilde{x}_i =  x_{i+1}-\Tilde{x}_{i+1}; \quad i = 1, \dots, n-1 \\
D^{\alpha}_{t}x_n-D^{\alpha}_{t}\Tilde{x}_n =  f(t,x)-f(t-T,\Tilde{x}) \\
\qquad \qquad \qquad \qquad + \; F^\top(t,x)\theta - F^\top(t-T,\Tilde{x})\Tilde{\theta} \\ 
\qquad \qquad \qquad \qquad + \; d - \Tilde{d} \\
\qquad \qquad \qquad \qquad + \; g(t,x)u(t) - g(t-T,\Tilde{x})\Tilde{u}(t).
\end{cases}
\end{align}

Let us use $e = x - \Tilde{x}$ to define the error vector. The equation \eqref{eq:error_1} can therefore be rewritten as
\begin{align}\label{eq:error_2}
\begin{cases}
D^{\alpha}_{t}e_i = e_{i+1}; \qquad 1\leq i \leq n-1\\
D^{\alpha}_{t}e_n = f(t,x) - f(t-T,\Tilde{x}) \\
\qquad\qquad + \; F^\top(t,x)\theta - F^\top(t-T,\Tilde{x})\Tilde{\theta}  \\
\qquad\qquad + \; d - \Tilde{d} + g(t,x)u(t) - g(t-T,\Tilde{x})\Tilde{u}(t).
\end{cases}
\end{align}

To stabilize a periodic orbit of the system, we should obtain the system input $u(t)$ such that the following requirements are fulfilled:
\begin{align}\label{eq:error_convergence}
    \lim_{t\to\infty}\norm{e(t)} = 0 \, \equiv \, \lim_{t\to\infty}\norm{x(t)-x(t-T)} = 0.
\end{align}

\begin{theorem}\label{thm:control_scheme}
Let $\Tilde{x}(t) = x(t-T)$ and $\Tilde{u}(t) = u(t-T)$, where $T$ is the period of the unstable periodic orbit of the chaotic system \eqref{eq:nonlinear_system}. If the following control input and adaptation laws are applied to the system \eqref{eq:nonlinear_system}, the chaotic behavior of the system is substituted by a regular periodic one.
\begin{gather}
	u = u_{eq} + u_{ad} + u_s, \label{eq:control_input_0} \\
	u_{eq} = -\frac{\sum_{i=1}^{n}\eta_i e_i + f(t,x) - f(t-T,\Tilde{x}) - g(t-T,\Tilde{x})\Tilde{u}}{g(t,x)}, \label{eq:control_input_1} \\
	u_{ad} = -\frac{F^\top(t,x)\hat{\theta}(t) - F^\top(t-T,\Tilde{x})\hat{\theta}(t-T) + 2\hat{k}\text{sgn}(S)}{g(t,x)}, \label{eq:control_input_2} \\
	u_{s} = -\frac{(M+\mu)\text{sgn}(S)}{g(t,x)}, \label{eq:control_input_3} \\
	D_{t}^{\alpha}\hat{\theta}_i = \gamma_i S \big(F_i(t,x) - F_i(t-T,\Tilde{x}) \big); \quad i=1,\dots,m,  \label{eq:adaptation_law_1} \\
	D_{t}^{\alpha}\hat{k} = 2\gamma_k \abs{S}, \label{eq:adaptation_law_2}
\end{gather}
where $\hat{\theta}_i\in\mathbb{R}$ and $\hat{k}\in\mathbb{R}$ are estimates of $\theta_i\in\mathbb{R}$ and $k\in\mathbb{R}$; $\theta_i\in\mathbb{R}$ and $F_i\in\mathbb{R}$ are elements of the vectors $\theta\in\mathbb{R}^m$ and $F\in\mathbb{R}^m$, respectively; $\mu\in\mathbb{R}$ is an arbitrary positive number; $\gamma_k\in\mathbb{R}$ and $\gamma_i\in\mathbb{R}$ ($i=1,\dots,m$) are adaptation coefficient; $\text{sgn}(\cdot)$ is the sign function; and $M\in\mathbb{R}$ is a sufficiently large positive constant. In the above equations, $M\in\mathbb{R}$ is a positive real constant number that is sufficiently large. In the Eq. \eqref{eq:adaptation_law_2}, $S\in\mathbb{R}$ is a sliding surface defined by
\begin{align}\label{eq:sliding_surface}
	S = e_n + \sum_{i=1}^{n}\eta_i I_{t}^{\alpha}e_i,
\end{align}
where $\eta_i$, $i=1,\dots,n$, are positive real constant numbers chosen such that the dynamic of the sliding surface is stabilized asymptotically, $S=0$, is provided.
\end{theorem}

\begin{remark}
All the system state variables and the sliding surface have to be continuously differentiable and measurable since this is a simplifying requirement that is frequently seen in controlling FO systems. This presumption is typical, particularly when a FO system is being attempted to be controlled. \cite{li2007remarks}. 
\end{remark}

\begin{remark}
For the existence of a sliding mode, it is necessary that the dynamics obtained by $S=0$ be asymptotically stable. By applying the time derivative of order $\alpha$ to equation \eqref{eq:sliding_surface}, we obtain
\begin{align}\label{eq:sliding_surface_der}
	D_t^{\alpha}S = D_t^{\alpha}e_n + \sum_{i=1}^{n}\eta_i e_i.
\end{align}

Therefore, the sliding surface dynamics can be obtained as follows
\begin{align}
	D_t^{\alpha}e_n = -\sum_{i=1}^{n}\eta_i e_i,
\end{align}
where $\eta_1, \dots, \eta_n$ are chosen such that the roots of equation $s^{n\alpha} + \sum_{i=1}^{n}\eta_i s^{(i-1)\alpha} = 0$ satisfy the condition \eqref{eq:nonlinear_system} and the sliding surface is asymptotically stable. Using Theorem \ref{thm:linear_system_stability}, this condition can be expressed as
\begin{align}
	\abs{\arg(s_i)}> \alpha\frac{\pi}{2},
\end{align}
where $s_i$, $i = 1,\cdots, n$, denote the roots of the above-mentioned equation.
\end{remark} 
\begin{proof}
Consider the following Lyapunov function 
\begin{align}\label{eq:lypunov_func}
	V = \frac{1}{2}S^2 + \sum_{i=1}^{m}\frac{1}{2\gamma_i}(\theta_i - \hat{\theta}_i)^2 + \frac{1}{2\gamma_k}(k-\hat{k})^2.
\end{align}

By applying the time derivative of order $\alpha$ to the Lyapunov function \eqref{eq:lypunov_func}, and using Lemma \ref{lemma:derivativeLF} along with some algebraic manipulations, we can obtain the following inequality for the derivative of the Lyapunov function:
\begin{align}\label{eq:lypunov_func_der_1}
D_t^{\alpha}V & \leq SD_t^{\alpha}S - \sum_{i=1}^{m}\frac{1}{\gamma_i}(\theta_i-\hat{\theta}_i)D_t^{\alpha}\hat{\theta}_i - \frac{1}{\gamma_k}(k-\hat{k})D_t^{\alpha}\hat{k} \nonumber\\
& \leq S\Big(D_t^{\alpha}e_n +\sum_{i=1}^{n}\eta_i e_i\Big) - \sum_{i=1}^{m}\frac{1}{\gamma_i}(\theta_i-\hat{\theta}_i)D_t^{\alpha}\hat{\theta}_i \nonumber \\
& \quad - \frac{1}{\gamma_k}(k-\hat{k})D_t^{\alpha}\hat{k}.
\end{align}
Substituting \eqref{eq:error_2} into \eqref{eq:lypunov_func_der_1} yields:
\begin{align}\label{eq:lypunov_func_der_2_2}
D_t^{\alpha}V \leq ~ & S\Big(\sum_{i=1}^{n}\eta_i e_i +f(t,x)-f(t-T,\tilde{x}) + F^\top(t,x)\theta \nonumber \\
& - F^\top(t-T,\tilde{x})\tilde{\theta} + d-\tilde{d} \nonumber \\
& + g(t,x)u-g(t-T,\tilde{x})\tilde{u} \Big) \nonumber   \\ 
& - \;\sum_{i=1}^{m}\frac{1}{\gamma_i}(\theta_i-\hat{\theta}_i)D_t^{\alpha}\hat{\theta}_i
\nonumber   \\
& - \frac{1}{\gamma_k}(k-\hat{k})D_t^{\alpha}\hat{k} .
\end{align}
Since it is assumed that the disturbance is bounded with an unknown upper bound, $\abs{d(t)}\leq k$, one may easily check that $\abs{d-\tilde{d}}\leq 2k$, and substituting this upper bound into Eq. \eqref{eq:lypunov_func_der_2_2} yields:
\begin{align}\label{eq:lypunov_func_der_3}
D_t^{\alpha}V \leq ~ & S\Big(\sum_{i=1}^{n}\eta_i e_i +f(t,x) - f(t-T,\tilde{x}) \nonumber \\
&  + F^\top(t,x)\theta - F^\top(t-T,\tilde{x})\tilde{\theta} \nonumber \\
&  + g(t,x)u-g(t-T,\tilde{x})\tilde{u} \Big) \nonumber  \\ 
& - \;\sum_{i=1}^{m}\frac{1}{\gamma_i}(\theta_i-\hat{\theta}_i)D_t^{\alpha}\hat{\theta}_i \nonumber \\
& - \frac{1}{\gamma_k}(k-\hat{k})D_t^{\alpha}\hat{k} + 2k\abs{S} .
\end{align}

Using control inputs \eqref{eq:control_input_0}-\eqref{eq:control_input_2} in inequality \eqref{eq:lypunov_func_der_3} yields:
\begin{align}\label{eq:lypunov_func_der_4}
D_t^{\alpha}V \leq & ~ S\Big(F^\top(t,x)\theta - F^\top(t-T,\tilde{x})\tilde{\theta} - F^\top(t,x)\hat{\theta} \nonumber \\
& + F^\top(t-T,\tilde{x})\hat{\theta} + g(t,x)u_s\Big)  \nonumber \\ 
& + \; 2k\abs{S} - 2\hat{k}\abs{S} - \sum_{i=1}^{m}\frac{1}{\gamma_i}(\theta_i-\hat{\theta}_i)D_t^{\alpha}\hat{\theta}_i \nonumber \\
& -\frac{1}{\gamma_k}(k-\hat{k})D_t^{\alpha}\hat{k}.
\end{align}

Adding the term $S( -F^\top(t-T,\tilde{x})\theta + F^\top(t-T,\tilde{x})\theta )$ to the right-hand side of the inequality \eqref{eq:lypunov_func_der_4} results in 
\begin{align}\label{eq:lypunov_func_der_5}
D_t^{\alpha}V \leq ~ & S\Big( F^\top(t,x)\theta - F^\top(t-T,\tilde{x})\theta \nonumber \\
& + F^\top(t-T,\tilde{x})\theta - F^\top(t-T,\tilde{x})\Tilde{\theta} \nonumber \\
& - F^\top(t,x)\hat{\theta} + F^\top(t-T,\tilde{x})\hat{\theta} + g(t,x)u_s\Big) \nonumber  \\
& + \; 2\abs{S}(k - \hat{k}) -\sum_{i=1}^{m}\frac{1}{\gamma_i}(\theta_i-\hat{\theta}_i)D_t^{\alpha}\hat{\theta}_i \nonumber \\
& - \frac{1}{\gamma_k}(k-\hat{k})D_t^{\alpha}\hat{k}.
\end{align}

After some mathematical manipulations, inequality \eqref{eq:lypunov_func_der_5} can be written as:
\begin{align}\label{eq:lypunov_func_der_6}
D_t^{\alpha}V \leq & ~ S\Big(F^\top(t-T,\tilde{x})(\theta - \tilde{\theta})+g(t,x)u_s\Big) \nonumber\\
& + (k-\hat{k})\Big(2\abs{S} - \frac{1}{\gamma_k}D_t^{\alpha}\hat{k}\Big)  \nonumber\\ 
& + \; \sum_{i=1}^{m}\bigg( (\theta_i-\hat{\theta}_i)\big( S(F_i(t,x)-F_i(t-T,\tilde{x}))\big) \bigg) \nonumber\\
& -\sum_{i=1}^{m}\bigg( \frac{1}{\gamma_i}(\theta_i-\hat{\theta}_i)D_t^{\alpha}\hat{\theta}_i \bigg) .
\end{align}
Substituting adaptation laws \eqref{eq:adaptation_law_1} and \eqref{eq:adaptation_law_2} into inequality \eqref{eq:lypunov_func_der_6} yields:
\begin{align}\label{eq:lypunov_func_der_7}
D_t^{\alpha}V \leq S\Big(F^\top(t-T,\tilde{x})\big(\theta(t) - \theta(t-T)\big)+g(t,x)u_s\Big). 
\end{align}

One can infer that $F^\top(t-T,\tilde{x})(\theta(t) - \theta(t-T))$ is bounded because $\theta$ is bounded. Using \eqref{eq:control_input_3} in inequality \eqref{eq:lypunov_func_der_7}, and considering that $M$ is a positive constant number that is sufficiently large, results in
\begin{align}\label{eq:lypunov_func_der_8}
D_t^{\alpha}V \leq -\mu \abs{S} .
\end{align}
Integrating both sides of \eqref{eq:lypunov_func_der_8}, we have
\begin{equation}\label{eq:lypunov_func_der_9}
\begin{array}{l}
I_t^{\alpha}D_t^{\alpha}V = V(t) - V(0) \leq - I_t^{\alpha}(\mu \abs{S}) = - \mu I_t^{\alpha}(\abs{S}) \\
\Rightarrow ~ V(t) + \mu I_t^{\alpha}(\abs{S}) \leq V(0) .
\end{array}
\end{equation}

The following equation can be obtained from the fact that the Lyapunov function $V(t)$ is positive definite:  
\begin{align}\label{eq:lypunov_func_der_10}
\mu I_t^{\alpha}(\abs{S}) \leq V(0) \quad \Rightarrow \quad I_t^{\alpha}(\abs{S}) \leq \frac{V(0)}{\mu} .
\end{align}

It is a uniformly continuous function since we presume that the sliding surface is continuously differentiable. Therefore, using Lemma \ref{lemma:barbalat}, inequality \eqref{eq:lypunov_func_der_10} indicates that as time tends toward infinity, the sliding surface decreases to zero. The error trajectory also converges to zero as a result of the origin's asymptotic stability. Thus, condition \eqref{eq:error_convergence} is satisfied, and the control objective is achieved. Therefore, Theorem \ref{thm:control_scheme} has been proved completely.
\end{proof}

\begin{remark}
As a special case, if $\theta(t) = \theta(t-T)$ or $\theta$ is constant, then $u_s$ can be selected as:
\begin{align}\label{eq:remark_4_3_1}
u_s = -\frac{\mu\; \mbox{sgn}(S)}{g(t,x)}.
\end{align}
\end{remark}

\begin{proof}
As we know $\theta(t) = \theta(t-T)$, so \eqref{eq:lypunov_func_der_7} will be simplified as follows:
\begin{align}\label{eq:remark_4_3_2}
D_t^{\alpha}V \leq S\,g(t,x)u_s.
\end{align}
By substituting \eqref{eq:remark_4_3_2} into \eqref{eq:remark_4_3_1}, one can obtain \eqref{eq:lypunov_func_der_8}.
\end{proof}

\begin{remark}
The implementation of control input may clatter as a result of the sign function's application in the equations of $u_{ad}$ and $u_s$.
Saturation or sigmoid functions can take the place of the sign function to avoid this problem. These functions are defined as follows \cite{slotine1991applied}
\begin{align}\label{eq:test1}
    \text{sat}\bigg(\frac{S}{\delta}\bigg)= 
    \begin{cases}
        {\frac{S}{\delta};} & {\abs{\frac{S}{\delta}}<1}\\
        {\text{sgn}\bigg(\frac{S}{\delta}\bigg);} & {\abs{\frac{S}{\delta}}\geq1}
    \end{cases},
\end{align}
where $\delta$ is a small positive real number.
\begin{align}
    \mbox{sigmoid}(S) = \frac{1}{1 + e^{-S}} \cdot
\end{align}
\end{remark}

\section{Simulation Results}\label{sec:simulation_results}
The proposed scheme has been evaluated on FO Duffing and Gyro systems to demonstrate its effectiveness in controlling chaotic behavior.

\subsection{FO Duffing System}
In this simulation, the well-known FO Duffing system is considered as follows:
\begin{align}\label{eq:duffing_1}
\begin{cases}
D_t^{\alpha}x_1 = x_2 \\
D_t^{\alpha}x_2 = - \; \theta_1x_1 - \theta_2x_1^3 - \theta_3x_2 + \theta_4 \cos(\omega t).
\end{cases}
\end{align}

By setting $\theta_1 = -1$, $\theta_2 = 1$, $\theta_3 = 0.15$, $\theta_4 = 0.3$, $\omega=1$, and $\alpha=0.96$, the FO Duffing equation shows chaotic behavior. Moreover, the existence of an unstable periodic orbit in the FO Duffing system with these parameter values is proven in the literature \cite{rahimi2012stabilizing}. Adding the external disturbance $d(t)$ and control input $u(t)$ to the system dynamics gives:
\begin{align}\label{eq:duffing_2}
\begin{cases}
D_t^{0.98}x_1 =  x_2 \\
D_t^{0.98}x_2 =  - \; \theta_1x_1 - \theta_2x_1^3 - \theta_3x_2 + \theta_4 \cos(\omega t)  \\
\qquad\qquad ~~ + d(t) + g(t,x)u(t).
\end{cases}
\end{align}

In this case, $d(t)$ is chosen as $0.2\cos(2t)$ which is bounded by $\abs{d(t)}<k=0.2$. The parameter $k$ is assumed to be unknown that should be updated by adaptation law \eqref{eq:adaptation_law_2}. To use the proposed method, we consider $f(t,x)=0$ and $g(t,x) = 1+x_1^2$. In addition, $F(t,x)$ and $\theta$ are defined as follows:

\begin{align}
F(t,x) = & \big[-x_1 \quad -x_1^3 \quad -x_2 \quad \cos(\omega t) \big]^\top, \\
\theta = & \big[\theta_1 \quad \theta_2 \quad \theta_3 \quad \theta_4 \big]^\top.
\end{align}
Therefore, \eqref{eq:duffing_2} can be rewritten as follows
\begin{align}
\begin{cases}
D_t^{0.98}x_1 = x_2 \\
D_t^{0.98}x_2 = f(t,x) + F^\top(t,x)\theta \\
\qquad \qquad ~~ + d(t) + g(t,x)u(t).
\end{cases}
\end{align}

The initial conditions are considered as $\hat{k}(0)=0.1$, $\theta(0) =[-1.5 \quad 1.5 \quad 0.2 \quad 0.5]^\top$, and $x(0) =[0.15 \quad 0.1]^\top$. The  adaptation coefficients are established to be $\gamma_k = 1$, and $\gamma =[5 \quad 5 \quad 5 \quad 5]^\top$. The Predictive-Evaluate-Correct-Evaluate method \cite{diethelm2005algorithms} is used to solve the fractional differential equations and the time step size is set to $0.005 s$. To resolve the fractional differential equations, the Predictive-Evaluate-Correct-Evaluate algorithm \cite{diethelm2005algorithms} is used with a time step size of $0.005 s$. 

\begin{figure}[!h]
\centering
\includegraphics[width=8.5cm]{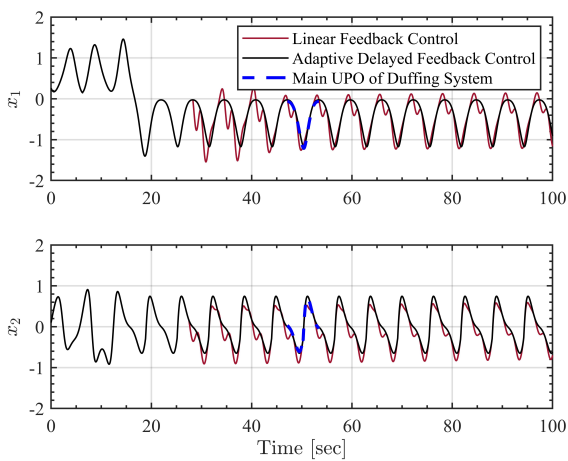}
\caption{Stabilization of the fractional Duffing system's UPO with T = 2$\pi$ via adaptive delayed feedback control and linear feedback control algorithms: a) $x_1$ time series, b) $x_2$ time series.}
\label{fig:duffing_state_variables}
\end{figure}

\begin{figure}[!h]
\centering 
\includegraphics[width=8.5cm]{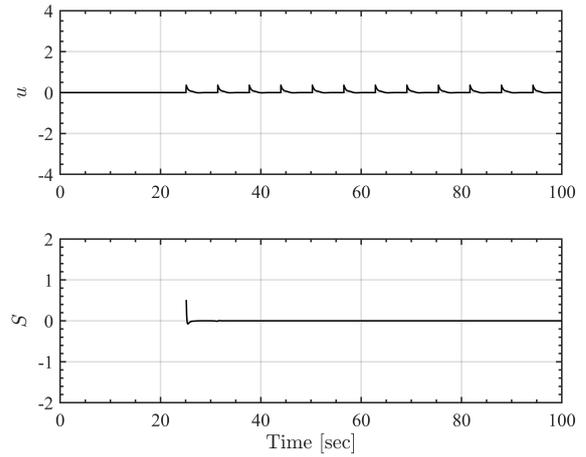}\\
\caption{(a) Trajectory of $u$, (b) Trajectory of $S$.}
\label{fig:duffing_control_sliding}
\end{figure}

The proposed adaptive delayed feedback control algorithm is applied to a well-known fractional Duffing system to stabilize its UPO with period $T = 2\pi$ in Figures \ref{fig:duffing_state_variables}-\ref{fig:duffing_control_sliding}. Figure \ref{fig:duffing_state_variables} (solid black line) illustrates the time history of the state variables and the main UPO of the system by the dashed blue lines. It is worth mentioning that the controller is applied at $t = 4T$. 
As these figures demonstrate, the UPO of the FO Duffing system is completely stabilized after almost $10$ seconds of applying the controller. So the chaotic behavior of the system is suppressed, and the state trajectories of the system track the main UPO. Although we assumed that $f = 0$ and all the system’s parameters are unknown, the controller can stabilize the system and this shows the adaptivity of the controller.  Additionally, the controller is able to stabilize the system's UPO regardless of the presence of uncertainty and external disturbance with unknown upper bounds, illustrating the proposed algorithm's robustness against system uncertainty and disturbance.  

Figure \ref{fig:duffing_control_sliding} illustrates the trajectory of the sliding surface and control input. As the system approaches the Unstable Periodic Orbit (UPO), a lower control signal is needed. The trajectory sets on the UPO with zero control signal when the chaotic system trajectory reaches the exact system UPO. However, due to unknown parameters and disturbances, the primary UPO may not be fully stabilized, and the controller action may not completely converge to zero. Consequently, the system  trajectories converge to the close vicinity of the main UPO.

\begin{figure}[!h]
\centering 
\includegraphics[width=8.5cm]{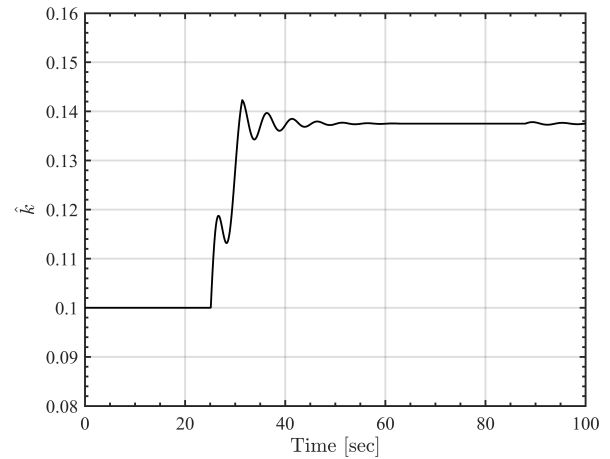}\\
\caption{Time series of the estimated upper bound of the external disturbance for the FO Duffing system}
\label{fig:duffing_k_hat}
\end{figure}

The time history for the estimated value of the external disturbance's upper bound in the FO Duffing system is illustrated in Figure \ref{fig:duffing_k_hat}. It can be observed that similar to the system's UPO, the estimated value represented by $\hat{k}$, converges to its final value after almost 30 seconds. Since the estimated value is the upper bound of the disturbance, even though it converges to the final value slower than other parameters, the UPO is stabilized before $\hat{k}$ converges to the final value.

\begin{figure}[!h]
\centering 
\includegraphics[width=8.5cm]{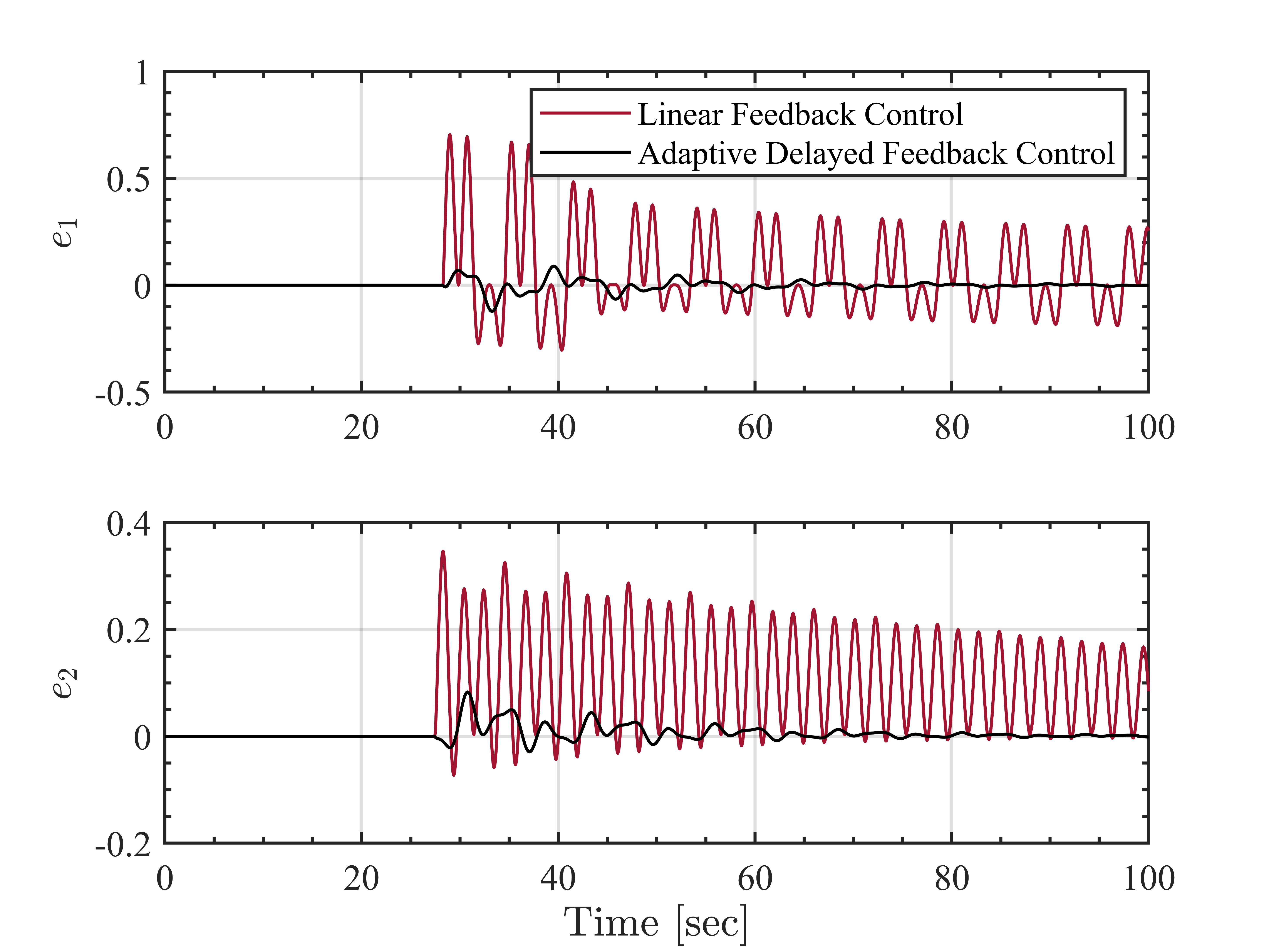}\\
\caption{UPO Tracking error of adaptive delayed feedback control and linear feedback control algorithms for the Duffing system. a) $e_1$ time series, b) $e_2$ time series.}
\label{fig:duffing_error}
\end{figure}

A linear feedback control method is developed for stabilizing the UPO of FO chaotic systems by \textit{Rahimi, et. al}\cite{rahimi2012stabilizing}.  We have applied the linear feedback control method to the duffing system and compared the results to our proposed algorithm. As Figure \ref{fig:duffing_state_variables} illustrates, our method converges to the main UPO of the system faster than the linear feedback method. Furthermore, Figure \ref{fig:duffing_error} compares the error of these two methods with the main system UPO. From the comparison, it can be concluded that the adaptive delayed feedback control scheme exhibits a lower steady-state error compared to the linear feedback control method.

\subsection{FO Gyro System}

In this subsection, we consider the FO chaotic Gyro system in the following form:

\begin{align}\label{eq:gyro_1}
\begin{cases}
D_t^{\alpha}x_1 = x_2 \\
D_t^{\alpha}x_2 = - \; \psi_1^2 \dfrac{(1-\cos x_1)^2}{\sin^3x_1} - \psi_2 x_2 - \psi_3 x_2^3 \\
\qquad\qquad + \; \beta \sin x_1 + f \sin \omega t \sin x_1. 
\end{cases}
\end{align}
By setting $\alpha = 0.98$, $\psi_1 = 10$, $\psi_2 = 0.5$, $\psi_3 = 0.05$, $\beta = 1$, $\omega = 25$, and $f = 35.5$, the FO Gyro system exhibits chaos. Furthermore, the existence of an unstable periodic orbit in this system is shown in \cite{aghababa2013rich}.

Adding the external disturbance $d(t)$ and control input $u(t)$ to the system dynamics gives:
\begin{align}\label{eq:gyro_2}
\begin{cases}
D_t^{0.97}x_1 =  x_2 \\
D_t^{0.97}x_2 =  - \; \psi_1^2 \dfrac{(1-\cos x_1)^2}{\sin^3x_1} - \psi_2 x_2 - \psi_3 x_2^3 \\ 
\qquad\qquad\quad + \;\beta \sin x_1  + f \sin \omega t \sin x_1 \\
\qquad\qquad\quad + \;d(t) + g(t,x)u(t).
\end{cases}
\end{align}

The disturbance is considered as $0.1\sin(t)$ whose upper bound is $k=0.1$.  It is presumed that the disturbance upper bound is not known and ought to be estimated by the adaption law \eqref{eq:adaptation_law_2}. Similar to the previous example, we consider $f(t,x)=f \sin \omega t \sin x_1$ and $g(t,x) = 1+x_1^2$. Also, $F(t,x)$ and $\theta$ are defined as follows:
\begin{align}\label{eq:gyro_F}
F(t,x) = & \big[-\frac{(1 - \cos x_1)^2}{\sin^3x_1} \quad -x_2 \quad -x_2^3 \quad \sin x_1 \big]^\top, \\
\theta =& \big[\psi_1^2 \quad \psi_2 \quad \psi_3 \quad \beta \big]^\top.
\end{align}

The initial conditions are considered as $\hat{k}(0)=0.1$, $\theta(0) =[-1.5 \quad 1.5 \quad 0.2 \quad 0.5]^\top$, and $x(0) =[0.15 \quad 0.1]^\top$. The adaptation parameters are has been chosen as $\gamma =[2 \quad 2 \quad 2 \quad 2]^\top$ and  $\gamma_k = 2$.

\begin{figure}[!h]
\centering
\includegraphics[width=8.5cm]{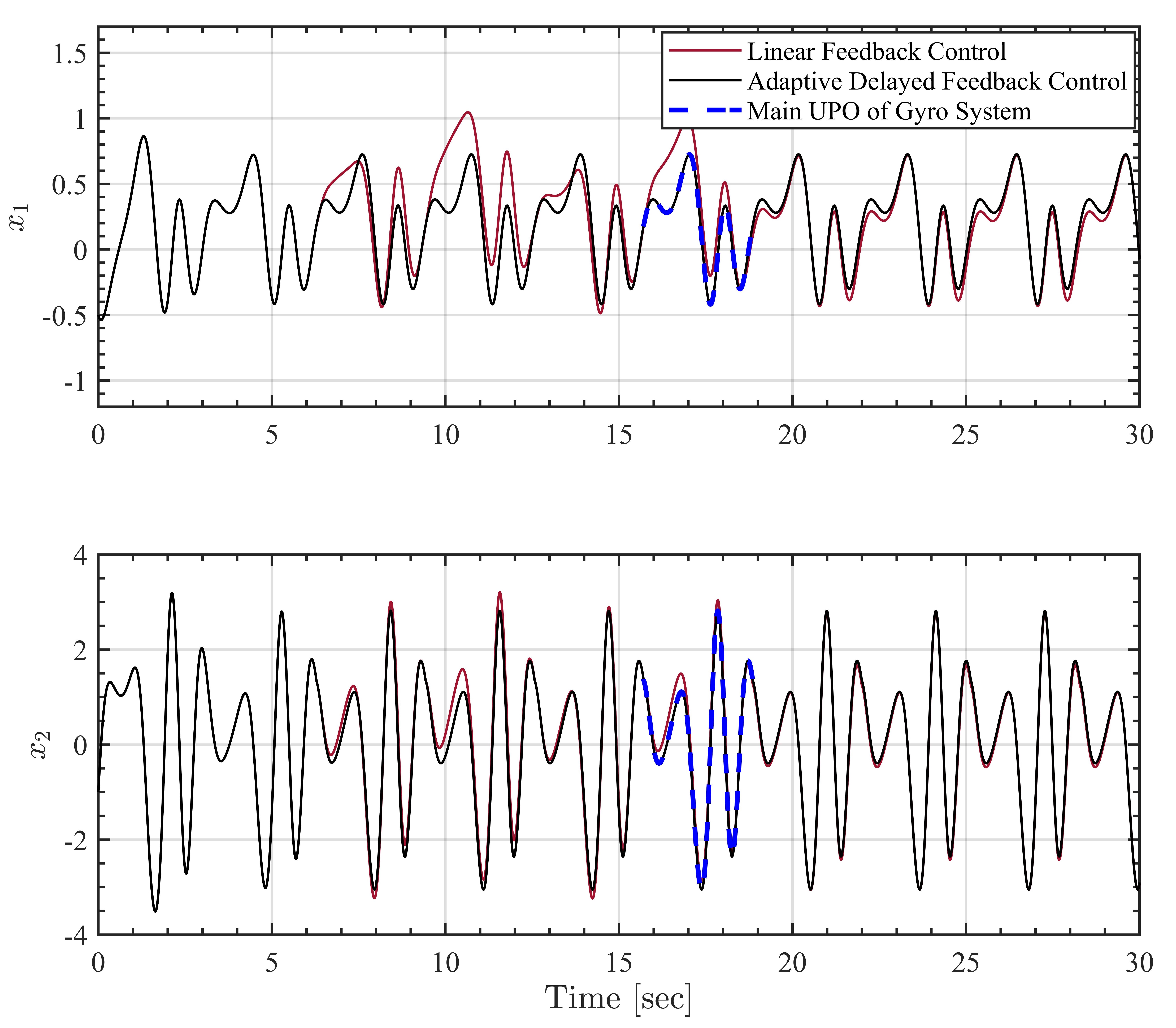}\\
\caption{Stabilization of the fractional Gyro system's UPO with T = 2$\pi$ via adaptive delayed feedback control and linear feedback control algorithms: a) $x_1$ time series, b) $x_2$ time series.}
\label{fig:gyro_state_variables}
\end{figure}

\begin{figure}[!h]
\centering 
\includegraphics[width=8.5cm]{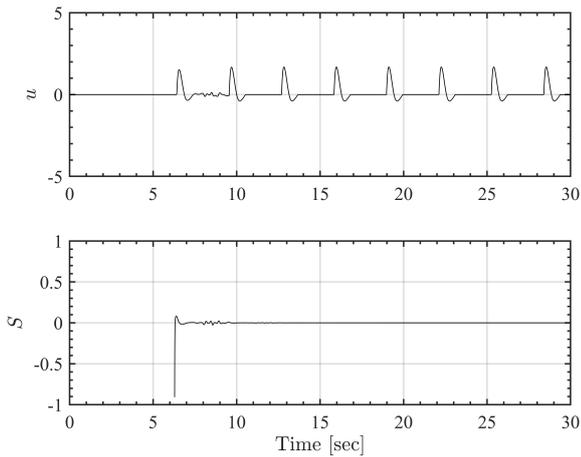}\\
\caption{(a) Trajectory of the control input $u$, (b) Trajectory of the sliding surface $S$.}
\label{fig:gyro_control_sliding}
\end{figure}

The simulation results are presented for stabilization of the periodic orbit of the Gyro system \eqref{eq:gyro_2} with period $T = 4\pi$ in Figures \ref{fig:gyro_state_variables}-\ref{fig:gyro_control_sliding}. Figure \ref{fig:gyro_state_variables} (solid black line) shows the time history of the state variables. In this figure, the main UPO of the system is demonstrated by the dashed blue lines. After $5$ seconds of applying the control input, the UPO is stabilized. Again similar to the previous part, to assess the robustness of the proposed control algorithm, system uncertainty and disturbance are added to the closed-loop system. Moreover, the adaptivity of the proposed algorithm is illustrated by assuming that all the system's parameters are unknown. Figure \ref{fig:gyro_control_sliding} demonstrates the time history of the control input and sliding surface. Since we have considered system uncertainty and disturbance in the numerical simulations, the trajectories of the system converge to close vicinity of the main UPO, and the control input, $u$, has not converged to zero.

\begin{figure}[!h]
\centering 
\includegraphics[width=8.5cm]{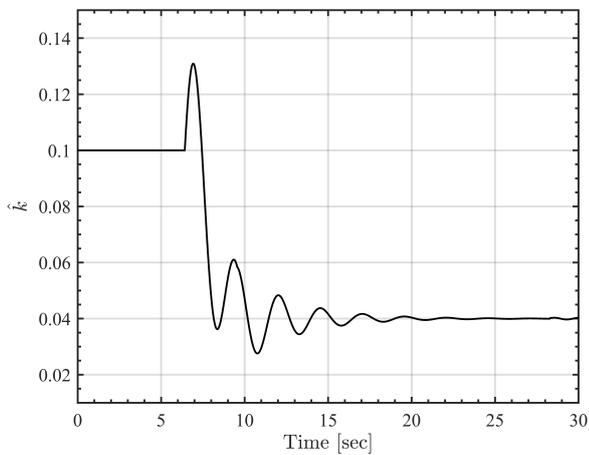}\\
\caption{Time series of the estimated upper bound of the external disturbance for the FO Gyro system.}
\label{fig:gyro_k_hat}
\end{figure}

Figure \ref{fig:gyro_k_hat} illustrates time history for the estimated value of the external disturbance's upper bound in system \eqref{eq:gyro_2}. After almost 16 seconds, $\hat{k}$ is convergent to its final value, much like the system's UPO. The upper bound of the disturbance may take longer to converge to its final value as compared to other parameters. However, despite this slower convergence rate, the system is able to stabilize on the UPO even before the estimated value reaches its final value.

We have applied the linear feedback control method to the duffing system and compared the results to our proposed algorithm. As Figure \ref{fig:duffing_state_variables} illustrates, our method converges to the main UPO of the system faster than the linear feedback method. Furthermore, Figure \ref{fig:duffing_error} compares the error of these two methods with the main system UPO. One can conclude that the steady-state error in the linear feedback control method is greater than the adaptive delayed feedback control scheme.

\begin{figure}[!h]
\centering 
\includegraphics[width=8.5cm]{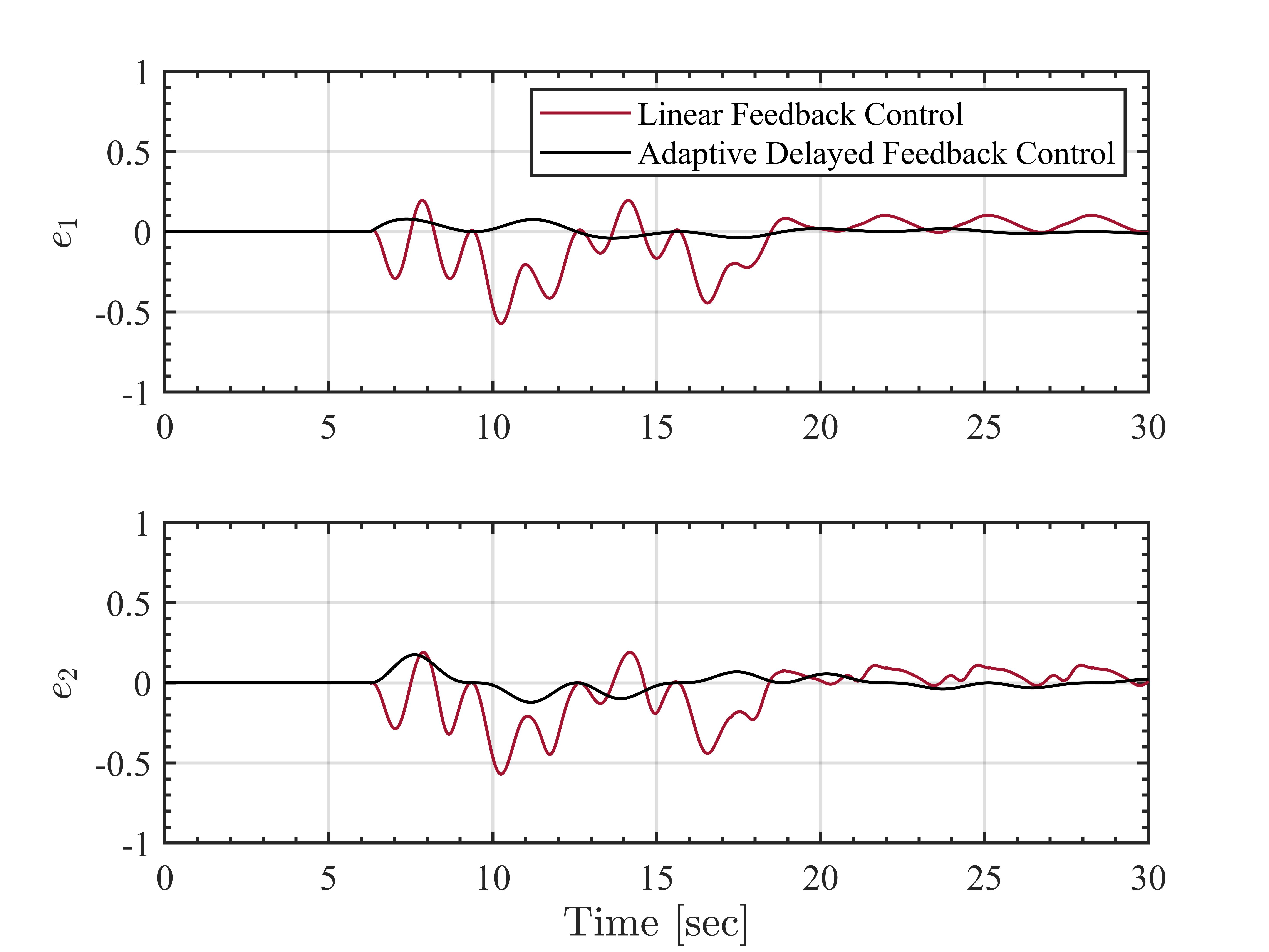}\\
\caption{UPO Tracking error of adaptive delayed feedback control and linear feedback control algorithms for the Gyro system. a) $e_1$ time series, b) $e_2$ time series.}
\label{fig:gyro_error}
\end{figure}

Similar to Duffing system, our presented method and the linear feedback control method have been applied to the fractional Gyro system. As depicted in Figure \ref{fig:gyro_state_variables} our algorithm demonstrates a higher convergence rate to the main system's UPO compared to the linear feedback control. Moreover, Figure \ref{fig:gyro_error} illustrates a comparison of the UPO tracking error between these two methods. Based on this comparison, it can be inferred that the adaptive delayed feedback control scheme exhibits a lower steady-state error than the linear feedback control technique.

\section{Conclusion}\label{sec:conclusion}
This article develops a robust adaptive nonlinear delayed feedback control strategy for a class of uncertain FO chaotic systems. The control framework employs the FO sliding mode control approach which guarantees the robustness of the closed-loop system against system uncertainties and external disturbances. The control input and adaptation mechanism are constructed from a proper sliding surface, using the Lyapunov approach. A key advantage of the proposed method is its independence from the UPO's trajectory, requiring only the period of an unstable periodic orbit for controller design. In the proposed framework, the stabilized orbit may not precisely represent the main UPO of the chaotic system due to external disturbances and system uncertainties. However, the stabilized orbit closely approximates the main UPO of the chaotic system. Finally, the proposed adaptive delayed feedback control scheme is applied to FO Duffing and Gyro systems, and numerical simulations are included to illustrate our findings and validate our theoretical results. These results show that the performance of our method is better than the linear feedback control method which is previously developed for stabilizing UPOs in FO chaotic systems in terms of steady-state error and convergence rate of tracking error.

Future research will focus on extending the proposed technique to systems with more general structures than system \eqref{eq:nonlinear_system} and evaluating the method in real-world systems through experimental setups. Additionally, future work will address the consideration of constraints on control inputs and state variables, and the extension of the Explicit Reference Governor (ERG) scheme to FO nonlinear systems. 

\section*{Declaration of conflicting interests}\label{sec:declarations}
The author(s) declared no potential conflicts of interest with respect to the research, authorship, and/or publication of this article.
\section*{Funding}\label{sec:Funding}
The author(s) received no financial support for the research, authorship, and/or publication of this article.
\section*{ORCID iDs}\label{sec:Funding}
Bahram Yaghooti \orcidlink{0000-0001-9646-9687} \href{https://orcid.org/0000-0001-9646-9687}{https://orcid.org/0000-0001-9646-9687} \\
Kaveh Safavigerdini \orcidlink{0000-0003-4904-0161} \href{https://orcid.org/0000-0003-4904-0161}{https://orcid.org/0000-0003-4904-0161} \\
Hassan Salarieh \orcidlink{0000-0002-0604-5731} \href{https://orcid.org/0000-0002-0604-5731}{https://orcid.org/0000-0002-0604-5731}

\bibliographystyle{SageV}
\bibliography{ref}

\begin{thebibliography}{10}
\providecommand{\url}[1]{\texttt{#1}}
\providecommand{\urlprefix}{URL }
\expandafter\ifx\csname urlstyle\endcsname\relax
  \providecommand{\doi}[1]{DOI:\discretionary{}{}{}#1}\else
  \providecommand{\doi}{DOI:\discretionary{}{}{}\begingroup
  \urlstyle{rm}\Url}\fi
\providecommand{\eprint}[2][]{\url{#2}}

\bibitem{mousa2020biohybrid}
Mousa MA, Soliman M, Saleh MA et~al.
\newblock Biohybrid soft robots, e-skin, and bioimpedance potential to build up
  their applications: A review.
\newblock \emph{IEEE Access} 2020; 8: 184524--184539.

\bibitem{azimirad2022consecutive}
Azimirad V, Ramezanlou MT, Sotubadi SV et~al.
\newblock A consecutive hybrid spiking-convolutional (chsc) neural controller
  for sequential decision making in robots.
\newblock \emph{Neurocomputing} 2022; 490: 319--336.

\bibitem{sahoo2019active}
Sahoo S and Ray M.
\newblock Active control of nonlinear transient vibration of laminated
  composite beams using triangular scld treatment with fractional order
  derivative viscoelastic model.
\newblock \emph{Journal of Dynamic Systems, Measurement, and Control} 2019;
  141(11).

\bibitem{engheta1996fractional}
Engheta N.
\newblock On fractional calculus and fractional multipoles in electromagnetism.
\newblock \emph{IEEE Transactions on Antennas and Propagation} 1996; 44(4):
  554--566.

\bibitem{shalalfeh2018modeling}
Shalalfeh L, Bogdan P and Jonckheere E.
\newblock Modeling of pmu data using arfima models.
\newblock In \emph{2018 Clemson University Power Systems Conference (PSC)}.
  IEEE, pp. 1--6.

\bibitem{yu2021extended}
Yu S, Feng Y and Yang X.
\newblock Extended state observer--based fractional order sliding-mode control
  of piezoelectric actuators.
\newblock \emph{Proceedings of the Institution of Mechanical Engineers, Part I:
  Journal of Systems and Control Engineering} 2021; 235(1): 39--51.

\bibitem{yaghooti2020constrained}
Yaghooti B, Hosseinzadeh M and Sinopoli B.
\newblock Constrained control of semilinear fractional-order systems:
  Application in drug delivery systems.
\newblock In \emph{2020 IEEE Conference on Control Technology and Applications
  (CCTA)}. IEEE, pp. 833--838.

\bibitem{yaghooti2023inferring}
Yaghooti B and Sinopoli B.
\newblock Inferring dynamics of discrete-time, fractional-order control-affine
  nonlinear systems.
\newblock In \emph{2023 American Control Conference (ACC)}. IEEE, pp. 935--940.

\bibitem{azar2017fractional}
Azar AT, Vaidyanathan S and Ouannas A.
\newblock \emph{Fractional order control and synchronization of chaotic
  systems}, volume 688.
\newblock Springer, 2017.

\bibitem{shahvali2022distributed}
Shahvali M, Naghibi-Sistani MB and Askari J.
\newblock Distributed adaptive dynamic event-based consensus control for
  nonlinear uncertain multi-agent systems.
\newblock \emph{Proceedings of the Institution of Mechanical Engineers, Part I:
  Journal of Systems and Control Engineering} 2022; 236(9): 1630--1648.

\bibitem{zamani2022formation}
Zamani H, Johari~Majd V and Khandani K.
\newblock Formation tracking control of fractional-order multi-agent systems
  with fixed-time convergence.
\newblock \emph{Proceedings of the Institution of Mechanical Engineers, Part I:
  Journal of Systems and Control Engineering} 2022; : 09596518221105788.

\bibitem{lori2021transportation}
Lori AAR, Danesh M, Amiri P et~al.
\newblock Transportation of an unknown cable-suspended payload by a quadrotor
  in windy environment under aerodynamics effects.
\newblock In \emph{2021 7th International Conference on Control,
  Instrumentation and Automation (ICCIA)}. IEEE, pp. 1--6.

\bibitem{amiri2020fuzzy}
Amiri P, Dayyani M, Lori AAR et~al.
\newblock Fuzzy--sliding mode versus integral sliding mode controller for a
  quadrotor with mass uncertainty under effect of wind.
\newblock In \emph{2020 28th Iranian conference on electrical engineering
  (ICEE)}. IEEE, pp. 1--7.

\bibitem{ayten2019implementation}
Ayten KK, {\c{C}}iplak MH and Dumlu A.
\newblock Implementation a fractional-order adaptive model-based pid-type
  sliding mode speed control for wheeled mobile robot.
\newblock \emph{Proceedings of the Institution of Mechanical Engineers, Part I:
  Journal of Systems and Control Engineering} 2019; 233(8): 1067--1084.

\bibitem{machado2017mathematical}
Machado JT and Lopes AM.
\newblock On the mathematical modeling of soccer dynamics.
\newblock \emph{Communications in Nonlinear Science and Numerical Simulation}
  2017; 53: 142--153.

\bibitem{shahvali2021adaptive}
Shahvali M, Naghibi-Sistani MB and Askari J.
\newblock Adaptive fault compensation control for nonlinear uncertain
  fractional-order systems: static and dynamic event generator approaches.
\newblock \emph{Journal of the Franklin Institute} 2021; 358(12): 6074--6100.

\bibitem{huong2020mixed}
Huong DC and Thuan MV.
\newblock Mixed $h_\infty$ and passive control for fractional-order nonlinear
  systems via lmi approach.
\newblock \emph{Acta Applicandae Mathematicae} 2020; 170(1): 37--52.

\bibitem{essa2017application}
Essa M, Aboelela MA and Hassan MAM.
\newblock Application of fractional order controllers on experimental and
  simulation model of hydraulic servo system.
\newblock In \emph{Fractional order control and synchronization of chaotic
  systems}. Springer, 2017.
\newblock pp. 277--324.

\bibitem{wang2019practical}
Wang Y, Li B, Yan F et~al.
\newblock Practical adaptive fractional-order nonsingular terminal sliding mode
  control for a cable-driven manipulator.
\newblock \emph{International Journal of Robust and Nonlinear Control} 2019;
  29(5): 1396--1417.

\bibitem{yaghooti2018robust}
Yaghooti B and Salarieh H.
\newblock Robust adaptive fractional order proportional integral derivative
  controller design for uncertain fractional order nonlinear systems using
  sliding mode control.
\newblock \emph{Proceedings of the Institution of Mechanical Engineers, Part I:
  Journal of Systems and Control Engineering} 2018; 232(5): 550--557.

\bibitem{askari2022sampling}
Askari I, Badnava B, Woodruff T et~al.
\newblock Sampling-based nonlinear mpc of neural network dynamics with
  application to autonomous vehicle motion planning.
\newblock In \emph{2022 American Control Conference (ACC)}. IEEE, pp.
  2084--2090.

\bibitem{askari2021nonlinear}
Askari I, Zeng S and Fang H.
\newblock Nonlinear model predictive control based on constraint-aware particle
  filtering/smoothing.
\newblock In \emph{2021 American Control Conference (ACC)}. IEEE, pp.
  3532--3537.

\bibitem{azimirad2020optimizing}
Azimirad V, Sotubadi SV and Sharifi FJ.
\newblock Optimizing the parameters of spiking neural networks for mobile robot
  implementation.
\newblock In \emph{2020 10th International Conference on Computer and Knowledge
  Engineering (ICCKE)}. IEEE, pp. 030--034.

\bibitem{nicotra2018explicit}
Nicotra MM and Garone E.
\newblock The explicit reference governor: A general framework for the
  closed-form control of constrained nonlinear systems.
\newblock \emph{IEEE Control Systems Magazine} 2018; 38(4): 89--107.

\bibitem{aghababa2013rich}
Aghababa MP and Aghababa HP.
\newblock The rich dynamics of fractional-order gyros applying a fractional
  controller.
\newblock \emph{Proceedings of the Institution of Mechanical Engineers, Part I:
  Journal of Systems and Control Engineering} 2013; 227(7): 588--601.

\bibitem{yaghooti2020adaptive}
Yaghooti B, Siahi~Shadbad A, Safavi K et~al.
\newblock Adaptive synchronization of uncertain fractional-order chaotic
  systems using sliding mode control techniques.
\newblock \emph{Proceedings of the Institution of Mechanical Engineers, Part I:
  Journal of Systems and Control Engineering} 2020; 234(1): 3--9.

\bibitem{farid2021finite}
Farid Y and Ruggiero F.
\newblock Finite-time disturbance reconstruction and robust fractional-order
  controller design for hybrid port-hamiltonian dynamics of biped robots.
\newblock \emph{Robotics and Autonomous Systems} 2021; 144: 103836.

\bibitem{vu2021iterative}
Vu M and Zeng S.
\newblock Iterative optimal control synthesis for nonlinear switching systems.
\newblock In \emph{2021 American Control Conference (ACC)}. IEEE, pp.
  998--1003.

\bibitem{oshima2019spatial}
Oshima K and Yanao T.
\newblock Spatial unstable periodic quasi-satellite orbits and their
  applications to spacecraft trajectories.
\newblock \emph{Celestial Mechanics and Dynamical Astronomy} 2019; 131: 1--32.

\bibitem{he2022homotopy}
He W, Huang Y, Wang J et~al.
\newblock Homotopy method for optimal motion planning with homotopy class
  constraints.
\newblock \emph{IEEE Control Systems Letters} 2022; 7: 1045--1050.

\bibitem{rahimi2012stabilizing}
Rahimi MA, Salarieh H and Alasty A.
\newblock Stabilizing periodic orbits of fractional order chaotic systems via
  linear feedback theory.
\newblock \emph{Applied Mathematical Modelling} 2012; 36(3): 863--877.

\bibitem{sadeghian2011control}
Sadeghian H, Salarieh H, Alasty A et~al.
\newblock On the control of chaos via fractional delayed feedback method.
\newblock \emph{Computers \& Mathematics with Applications} 2011; 62(3):
  1482--1491.

\bibitem{naceri2008prediction}
Naceri A, Mansouri N and Charef A.
\newblock Prediction-based feedback control of fractional order system.
\newblock In \emph{2008 IEEE International Symposium on Industrial
  Electronics}. IEEE, pp. 908--912.

\bibitem{zheng2015fuzzy}
Zheng Y.
\newblock Fuzzy prediction-based feedback control of fractional order chaotic
  systems.
\newblock \emph{Optik} 2015; 126(24): 5645--5649.

\bibitem{9615365}
Shahvali M, Naghibi-Sistani MB and Askari J.
\newblock Dynamic event-triggered control for a class of nonlinear
  fractional-order systems.
\newblock \emph{IEEE Transactions on Circuits and Systems II: Express Briefs}
  2022; 69(4): 2131--2135.
\newblock \doi{10.1109/TCSII.2021.3128561}.

\bibitem{rabah2015state}
Rabah K, Ladaci S and Lashab M.
\newblock State feedback with fractional pi$\lambda$d$\mu$ control structure
  for genesio-tesi chaos stabilization.
\newblock In \emph{2015 16th International Conference on Sciences and
  Techniques of Automatic Control and Computer Engineering (STA)}. IEEE, pp.
  328--333.

\bibitem{danca2017impulsive}
Danca MF, Fe{\v{c}}kan M and Chen G.
\newblock Impulsive stabilization of chaos in fractional-order systems.
\newblock \emph{Nonlinear Dynamics} 2017; 89(3): 1889--1903.

\bibitem{layeghi2008stabilizing}
Layeghi H, Arjmand MT, Salarieh H et~al.
\newblock Stabilizing periodic orbits of chaotic systems using fuzzy adaptive
  sliding mode control.
\newblock \emph{Chaos, Solitons \& Fractals} 2008; 37(4): 1125--1135.

\bibitem{podlubny1998fractional}
Podlubny I.
\newblock \emph{Fractional differential equations: an introduction to
  fractional derivatives, fractional differential equations, to methods of
  their solution and some of their applications}.
\newblock Elsevier, 1998.

\bibitem{matignon1996stability}
Matignon D.
\newblock Stability results for fractional differential equations with
  applications to control processing.
\newblock In \emph{Computational engineering in systems applications},
  volume~2. Lille, France, pp. 963--968.

\bibitem{li2010stability}
Li Y, Chen Y and Podlubny I.
\newblock Stability of fractional-order nonlinear dynamic systems: Lyapunov
  direct method and generalized mittag--leffler stability.
\newblock \emph{Computers \& Mathematics with Applications} 2010; 59(5):
  1810--1821.

\bibitem{duarte2015using}
Duarte-Mermoud MA, Aguila-Camacho N, Gallegos JA et~al.
\newblock Using general quadratic lyapunov functions to prove lyapunov uniform
  stability for fractional order systems.
\newblock \emph{Communications in Nonlinear Science and Numerical Simulation}
  2015; 22(1-3): 650--659.

\bibitem{zhang2017new}
Zhang R and Liu Y.
\newblock A new barbalat's lemma and lyapunov stability theorem for fractional
  order systems.
\newblock In \emph{2017 29th Chinese control and decision conference (CCDC)}.
  IEEE, pp. 3676--3681.

\bibitem{li2007remarks}
Li C and Deng W.
\newblock Remarks on fractional derivatives.
\newblock \emph{Applied Mathematics and Computation} 2007; 187(2): 777--784.

\bibitem{slotine1991applied}
Slotine JJE, Li W et~al.
\newblock \emph{Applied nonlinear control}, volume 199.
\newblock Prentice hall Englewood Cliffs, NJ, 1991.

\bibitem{diethelm2005algorithms}
Diethelm K, Ford NJ, Freed AD et~al.
\newblock Algorithms for the fractional calculus: a selection of numerical
  methods.
\newblock \emph{Computer methods in applied mechanics and engineering} 2005;
  194(6-8): 743--773.

\end{thebibliography}

\end{document}